\documentclass[twocolumn,apj,twocolappendix]{aastex631}
\usepackage{amsmath,amsfonts,amssymb,bm,bbm}
\usepackage[colorlinks=true]{hyperref}
\usepackage{xcolor}
\usepackage{graphbox}

\newcommand{\DLinkGNN}{10}
\newcommand{\DLinkOverdensity}{3}

\newcommand{\EBMbase}{\textsf{EBM-base}}
\newcommand{\EBMdisperse}{\textsf{EBM-DisPerSE}}
\newcommand{\EBMoverdensity}{\textsf{EBM-overdensity}}
\newcommand{\EBMall}{\textsf{EBM-all}}

\shorttitle{How the Galaxy--Halo Connection Depends on Large-Scale Environment}
\shortauthors{J.~F.~Wu et al.}

\begin{document} 

\title{How the Galaxy--Halo Connection Depends on Large-Scale Environment}

\author[0000-0002-5077-881X]{John F. Wu}
\affiliation{Space Telescope Science Institute, 3700 San Martin Dr, Baltimore, MD 21218}
\affiliation{Department of Physics \& Astronomy, Johns Hopkins University, 3400 N Charles St, Baltimore, MD 21218}
\email{jowu@stsci.edu}

\author[0000-0002-8896-6496]{Christian Kragh Jespersen}
\affiliation{Department of Astrophysical Sciences, Princeton University, Princeton, NJ 08544, USA}

\author[0000-0003-2229-011X]{Risa~H.~Wechsler}
\affiliation{Kavli Institute for Particle Astrophysics and Cosmology and Department of Physics, 452 Lomita Mall, Stanford University, Stanford, CA 94305, USA}
\affiliation{SLAC National Accelerator Laboratory, 2575 Sand Hill Road, Menlo Park, CA 94025, USA}

\begin{abstract}
We investigate the connection between galaxies, dark matter halos, and their large-scale environments at $z=0$ with Illustris TNG300 hydrodynamic simulation data.
We predict stellar masses from subhalo properties to test two types of machine learning (ML) models: Explainable Boosting Machines (EBMs) with simple galaxy environment features and $\mathbb{E}(3)$-invariant graph neural networks (GNNs).
The best-performing EBM models leverage spherically averaged overdensity features on $3$~Mpc scales. 
Interpretations via SHapley Additive exPlanations (SHAP) also suggest that, in the context of the TNG300 galaxy--halo connection, simple spherical overdensity on $\sim 3$~Mpc scales is more important than cosmic web distance features measured using the \texttt{DisPerSE} algorithm.
Meanwhile, a GNN with connectivity defined by a fixed linking length, $L$, outperforms the EBM models by a significant margin.
As we increase the linking length scale, GNNs learn important environmental contributions up to the largest scales we probe ($L = 10$~Mpc).
We conclude that $3$~Mpc distance scales are most critical for describing the TNG galaxy--halo connection using the spherical overdensity parameterization but that information on larger scales, which is not captured by simple environmental parameters or cosmic web features, can further augment these models.
Our study highlights the benefits of using interpretable ML algorithms to explain models of astrophysical phenomena, and the power of using GNNs to flexibly learn complex relationships directly from data while imposing constraints from physical symmetries.
\end{abstract}

\section{Introduction}\label{sec:intro}  

Galaxies form and co-evolve in tandem with their dark matter halos over cosmic timescales.
Although this galaxy--halo connection has been characterized using detailed cosmological hydrodynamic simulations, many important trends can be described via simple relationships and models.
For example, a galaxy's stellar mass $M_\star$ scales with its dark matter halo mass $M_{\rm halo}$, and this stellar-to-halo mass relation (SHMR) depends on whether the galaxy resides in a central or satellite subhalo.

The paper addresses two questions about large-scale galaxy environment and its impact on the galaxy--halo connection.
First, what is the optimal length for quantifying large-scale environment? 
Second, can a graph neural network (GNN) outperform ML models that are provided spherical overdensity and/or cosmic web features, thereby showing that these summary statistics are incomplete?
We will answer these questions by measuring how well various algorithms can estimate baryons purely from dark matter halo properties, i.e., recover the galaxy--halo connection for a single hydrodynamic simulation.

One way to model the galaxy--halo connection is via subhalo abundance matching (SHAM), a non-parametric mapping between stellar mass and subhalo mass \citep[e.g.,][]{ValeOstriker04}. 
The galaxy--halo connection can also be formulated using other kinds of models, such as halo occupation distribution HOD or conditional luminosity function (CLF) approaches \citep[e.g.,][]{PeacockSmith00,Seljak00,Benson+00,Berlind+02,Yang+03,Kravtsov+04,Zheng+05}.
For a comprehensive overview of the galaxy--halo connection, and outstanding questions, see the review by \citet{2018ARA&A..56..435W}.

We investigate the galaxy--halo connection through machine learning (ML) models that can exploit and quantify the impacts of large-scale galaxy environment at low redshift.
Previous studies have shown that ML models trained on hydrodynamic simulation outputs can more accurately populate baryonic properties onto dark matter subhalos than SHAM or HOD models \citep[e.g.,][]{2016MNRAS.457.1162K,2018MNRAS.478.3410A,2019MNRAS.490.2367C,2019arXiv190205965Z,Moster+2021,2022ApJ...941..132M,WuJespersen2023,Chuang+23_IllustrisMangrove}.

Galaxy physical properties and clustering properties depend on variables beyond halo mass, a phenomenon known as assembly bias.
At given halo mass, halos that formed earlier or have higher mass concentrations tend to cluster more strongly and are more often populated with central galaxies \citep[e.g.,][]{Wechsler+2002,ShethTormen2004,Wechsler+2006,Wang+2007,Hahn+2009,Mao+2018,Zehavi+2018}.
Incorporating a subhalo's merger history is helpful for predicting baryons from dark matter subhalos. 
For example, the galaxy--halo connection can be augmented by tracking satellite properties at time of accretion  \citep{Conroy+06,Christensen+2023}, by including merger history parameters \citep{Chuang23_abundancematching,Hausen+2023}, or by directly using the merger tree \citep{Jespersen+2022, Chuang+23_IllustrisMangrove}.

Assembly bias is sometimes also studied through the lens of galaxy environment.
The importance of galaxy environment beyond a central galaxy's virial radius has been debated \citep[e.g.,][]{Kauffmann+2004,Blanton+2006,Weinmann+2006}, but observational evidence for large-scale correlations in galaxy properties has been mounting with deeper and wider galaxy surveys.
While observations have revealed that redder galaxies preferentially populate overdense environments \citep[e.g.,][]{Dressler1980,Hogg+2003,Blanton+2005}, the physical processes that produce such correlations beyond galaxies' virial radii are not yet well understood \citep[e.g.,][]{2018ARA&A..56..435W,Zehavi+2018}.
For example, galaxy colors and star formation rates appear to be synchronized across $\sim 3$ Mpc scales \citep[``two-halo'' galactic conformity, see, e.g.,][]{Kauffmann+2013,Olsen+2021,OlsenGawiser2023}.
Galactic conformity has also been reproduced in simulations and semi-analytic models, although these studies sometimes disagree about the physical processes responsible for the signal \citep{Somerville08_SCSAM, Hearin+2015,Hearin+2016,Lacerna+2018,Lacerna+2022,Ayromlou+2023}.

There are several ways to parameterize a galaxy's surroundings.
One of the simplest methods is to define an average overdensity, $\delta$. 
Variations of this parameter can measure the total mass or galaxy number density within a specified radius, $L$ \citep[e.g.,][]{Blanton+2006,Tinker+2008}.
This definition of overdensity is equivalent to a convolution with a spherical ``top-hat'' filter over the mass or number counts.
Other works measure overdensity using the distance to the Nth nearest neighbor \citep[e.g.,][]{Peng+2010,Woo+2013}.

A constant length scale may not robustly characterize galaxy environment; galaxy large-scale structure is dynamic and cannot be fully captured by the positional configuration of galaxies in a single cosmic snapshot.
Therefore, detecting the \textit{persistent} cosmic web structures to characterize galaxy environments is crucial.
One of the most well-known methods, \texttt{DisPerSE}, relies on topological data analysis to identify filaments and nodes in the cosmic web \citep[see, e.g.,][]{Sousbie2011,Galarraga-Espinosa+2020,Hasan+2024,Wang+2023}.
The persistent cosmic web can be essential for understanding large-scale structure beyond mass distributions at a single snapshot; for example, galaxy spins are aligned with the cosmic web in simulations \citep[e.g.,][]{Pichon+2011,Laigle+2015}.

We focus on modeling the $z = 0$ relationship between galaxy stellar mass and dark matter subhalo properties because it is an important theoretical prediction from $\Lambda$CDM.
In particular, the SHMR is sensitive to many physical processes at both the low- and high-mass ends and has been well-calibrated in hydrodynamic simulations \citep[e.g.,][]{SomervilleDave2015,BullockBolyan-Kolchin2017}.
Moreover, a galaxy's stellar mass assembly also depends on physics that occurs over very different timescales \citep[see, e.g.,][]{Iyer+2020}, which must be captured in the ML models even though we do not directly account for the assembly history.
Therefore, we aim to understand how galaxy environment impacts a ML model's ability to estimate present-day stellar mass from $z=0$ dark matter subhalo catalogs.

The layout of our paper is as follows.
In Section~\ref{sec:data}, we describe the TNG300 simulation data used in this work.
In Section~\ref{sec:method}, we introduce the algorithms in our analysis, including Explainable Boosting Machines and Graph Neural Networks.
We present our results in Section~\ref{sec:results}, and interpret the models in Section~\ref{sec:interpretation}. 
We discuss the findings in Section~\ref{sec:discussion}.
Finally, we present our conclusions in Section~\ref{sec:conclusions}.
Several methodological details, model results, and discussions can be found in Appendix~\ref{app:GNN-details}.
We use the \cite{Planck2015} cosmology with $H_0 = 67.74\rm \ km\ s^{-1}\ Mpc^{-1}$ following the TNG300 simulations.
Throughout this work, we use base-10 logarithms.

\section{Data} \label{sec:data}

Our data set is based on \texttt{SUBFIND} subhalo catalogs \citep{Springel+2001} from the Illustris TNG300-1 cosmological hydrodynamic simulations \citep[hereafter, TNG300,][]{Nelson+2019b,Pillepich+2019}.\footnote{Additional details for these catalogs can be found online at \url{https://www.tng-project.org/data/docs/specifications/}.} 
Here, a subhalo is defined as a central halo or a satellite halo that may reside in a larger halo.
We select subhalos from TNG300 matched to their counterpart subhalos from the TNG300-1-Dark (dark matter only) simulation.
For our catalogs, subhalos must be detected by both the \texttt{Sublink} \citep{Sublink} and \texttt{LHaloTree}  \citep{LHaloTree} algorithms, which ensures a high level of confidence that the subhalos are well-resolved and matched correctly (but may result in some incompleteness).
By imposing this matching criterion, we remove 4.5\% of subhalos from the catalog.

We define $M_{\rm halo}$ as the sum of all dark matter particles gravitationally bound to the halo. 
$V_{\rm max}$ is the maximum value of the spherically averaged rotation curve. 
Both $M_{\rm halo}$ and $V_{\rm max}$ are extracted from the dark matter only subhalo catalogs.
The stellar mass $M_\star$ is defined using the sum of all star particles bound to the halo, based on the crossmatched hydrodynamic simulation.

We remove all flagged subhalos (\texttt{SubhaloFlag != 0}), as well as all subhalos with fewer than 50 star particles, halo masses below $\log(M_{\rm halo}/M_\odot) = 11$, or stellar masses below $\log (M_\star/M_\odot) = 9$.
For each subhalo, we use the $3d$ positions $\bm{x}$, $3d$ velocities $\bm{v}$, logarithmic halo mass $M_{\rm halo}$, maximum circular velocity $V_{\rm max}$, and whether the subhalo is a central or satellite, $\Gamma_{\rm cen} \in \{0,1\}$.
After applying these cuts, 76.1\% of the 208,037 remaining subhalos are centrals (in the dark matter only simulation).

We compute the total logarithmic halo mass within a radius of length $L$ for each subhalo, which we define as the spherical local overdensity $\delta_L$.\footnote{This is equivalent to convolving with a spherical top-hat filter with radius $L$.} 
We also augment the subhalo catalog with a table of cosmic web features generated for the TNG300 simulations \citep{Duckworth+2019} using the \texttt{DisPerSE} code \citep[][]{Sousbie2011}.
Specifically, we employ their catalog of distances to the nearest void, 1-saddle point, 2-saddle point, node, and skeleton.
In astronomy terminology, 1-saddle points are also known as walls, 2-saddle points are filaments, nodes are clusters or groups, and the skeleton is simply any of the above cosmic web features.
These cosmic web parameters are discussed further in Section~\ref{sssec:cosmic_web}.

\section{Methodology} \label{sec:method}

To probe the impact of galaxy environment on the modeled galaxy--halo connection, we rely on Explainable Boosting Machines (EBMs; Section~\ref{ssec:ebm}) and GNNs (Section~\ref{ssec:gnns}). 
We train a base model, \EBMbase{}, which takes as input the halo mass, $M_{\rm halo}$, maximum circular velocity, $V_{\rm max}$, and whether or not the subhalo is a central or satellite, $\Gamma_{\rm cen}$.
We train another model, \EBMdisperse{}, which extends the base model inputs using five features corresponding to distances from persistent cosmic web structures measured using the \texttt{DisPerSE} code.
Finally, we train another set of models, \EBMoverdensity{}, which consists of the base model augmented with subhalo mass overdensities, $\delta_L$, averaged over a spherical volume with radius $L$~Mpc.\footnote{Note the slight abuse of notation: we use $L$ to define distance scales for both the spherically averaged overdensity, as well as the graph connectivity later. These definitions are physically distinct. However, as we will see, both methods arrive at similar characteristic length scales, so we use $L$ in both contexts.}
We test values of $L \in \{0.3, 0.5, 1, 1.5, 2, 2.5, 3, 3.5, 4, 5, 7.5, 10\}$~Mpc.
For the GNN models, we test the same set of linking lengths used to connect subhalos during the construction of the cosmic graph. 
The GNN learns an optimal combination of flexible summary statistics to represent environmental information on scales up to $L$.

Readers who wish to skip over the details of our methodology may still wish to read about the baseline EBM model (Section~\ref{sssec:ebm-base}), the DisPerSE cosmic web features (Section~\ref{sssec:cosmic_web}), and the spherical overdensity features (\ref{sssec:overdensity}).
We also present an overview of the GNN inputs and model architecture in Figure~\ref{fig:gnn-schematic}.
Additional GNN details are supplied in Appendix~\ref{app:GNN-details}.

\subsection{Explainable Boosting Machines} \label{ssec:ebm}

Before we introduce EBMs, we begin by describing simpler interpretable models.
First, let us consider \textit{generalized linear models}:
\begin{equation}
    y = \beta_0 + \sum_{n=1}^N\beta_n x_n,
\end{equation} 
where the learnable model coefficients $\beta_1, \cdots, \beta_n$ are linear with the input features $x_1, \cdots, x_n$.
The model is interpretable because these coefficients can be regarded as feature importances (if the inputs span the same domain).
However, generalized linear models are often not expressive enough to fit complex relations between the independent and dependent variables when the underlying functional form is unknown.
This results in inaccurate predictions and erroneous interpretations of the best-fit coefficients.

We can extend generalized linear models by making predictions using non-linear functions, $f_n(x_n)$:
\begin{equation}
    y = \beta_0 + \sum_{i=n}^N f_n(x_n).
\end{equation}
These \textit{generalized additive models} \citep[][]{GAMs} can more flexibly fit combinations of input features.
In practice, splines or other basis functions can be used for the functions $f_n$.

Generalized additive models can be extended even further by allowing pairwise interactions \citep[][]{Lou+13}:  
\begin{equation}\label{eq:GAAM}
    y = \beta_0 + \sum_{n=1}^N f_n(x_n) + \sum_{n=1}^{N}\sum_{m=n+1}^N f_{mn}(x_m, x_n), 
\end{equation}
These generalized additive models with pairwise interactions can not only offer interpretable model components but also produce competitive results compared to other popular ML algorithms such as random forests \citep{randomforest}, gradient boosted trees \citep{gradientboosting}, or neural networks \citep[e.g.,][]{LeCun+1989}.

EBMs are generalized additive models with pairwise interactions, where the non-linear functions are binned look-up tables.
EBMs are also optimized using a specific routine; to train them, we iteratively build the model described in Equation~\ref{eq:GAAM} one feature at a time and minimize the residuals at each step (i.e. gradient boosting).
We implement and train the EBM using the \texttt{ExplainableBoostingRegressor} class from the \texttt{InterpretML} Python module \citep{InterpretML}.
The model has no trouble converging on a solution to fit stellar masses from halo properties.
To maximize performance, we set the number of bins per feature to 50,000 and the number of interactions to 32, which we found to be effective via a quick hyperparameter scan.

\subsubsection{The \EBMbase{} model} \label{sssec:ebm-base}

The aim of the \EBMbase{} model is to provide a strong baseline estimate of the galaxy--halo relation using only simple summary statistics.
Since the EBM models use one- and two-dimensional look-up tables for the $f_n$ and $f_{mn}$ functions in Equation~\ref{eq:GAAM}, respectively, they are similar to conditional abundance matching models.
The \EBMbase{} can learn univariate and pair-wise interactions between the input variables $M_{\rm halo}$, $V_{\rm max}$, and $\Gamma_{\rm cen}$.

\subsubsection{The \EBMdisperse{} model} \label{sssec:cosmic_web}

We also include cosmic web features based on discrete Morse and persistence theories that quantify the large-scale topology \citep[e.g.,][]{Pogosyan+2009,Pichon+2011}, produced from the \texttt{DisPerSE} code \citep[][]{Sousbie2011}.
Persistence quantifies the robustness of topological structures to local perturbations. 
In the context of galaxies and large-scale structure, higher persistence results in a more reliable cosmic web skeleton while filtering out small-scale features.
Significant work has shown that \texttt{DisPerSE} is essential for extracting and characterizing the cosmic web in both simulated and observed data \citep[e.g.,][]{Bonjean+2020,Galarraga-Espinosa+2020,Malavasi+2020,Rost+2021,Galarraga-Espinosa+2022,Hasan+2024,Hoosain+2024}.

We use five \texttt{DisPerSE} features that have been run on the TNG300 catalogs and published online:\footnote{A persistence of $4\sigma$ was used to generate these catalogs. For more details, see the code here: \url{https://github.com/illustristng/disperse_TNG}.} d\_minima, d\_saddle\_1, d\_saddle\_2, d\_node, d\_skel \citep[][see also Section~\ref{sec:data}]{Duckworth+2019,Duckworth+2020}.
These features quantify each subhalo's position relative to persistent large-scale structures such as voids, walls, filaments, clusters, and the nearest portion of the cosmic web skeleton, respectively.
Given \texttt{DisPerSE}'s ability to encode information about the cosmic web, we expect these features to be critical for learning an improved galaxy--halo relation.
We train the \EBMdisperse{} model to predict stellar masses using the base subhalo features ($M_{\rm halo}$, $V_{\rm max}$, $\Gamma_{\rm cen}$) supplemented by the five cosmic web distance features.

\subsubsection{The \EBMoverdensity{} models} \label{sssec:overdensity}

For the \EBMoverdensity{} model, we augment the base subhalo features ($M_{\rm halo}$, $V_{\rm max}$, $\Gamma_{\rm cen}$) with the averaged spherical overdensity on $L$ scales, $\delta_L$.
Here, the spherical overdensity is computed as the summed mass of all subhalos within a radius of $L$ Mpc, which serves as a simple measure of the large-scale environment
\citep{Blanton+2006,Tinker+2008}.
We retrain and cross-validate \EBMoverdensity{} models while varying $L \in \{0.3, 0.5, 1, 1.5, 2, 2.5, 3, 3.5, 4, 5, 7.5, 10\}$~Mpc to probe a wide range of length scales.
Length scales larger than $10$~Mpc are poorly sampled in TNG300, so we restrict our analysis to the range of galaxy environments on 0.3 to 10 Mpc length scales.

\subsection{Graph Neural Networks} \label{ssec:gnns}

\begin{figure*}[ht]
    \centering
    \includegraphics[width=\textwidth]{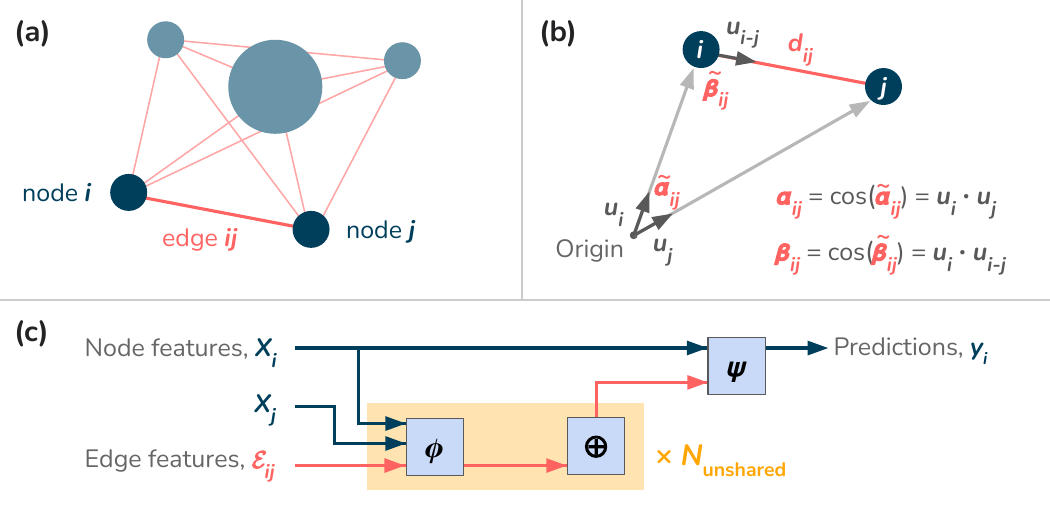}
    \caption{
    \textbf{A schematic showing the graph features and GNN model.} Nodes and node features are shown in blue, while edges and edge features are shown in red. (a) We depict a toy example of five subhalos (nodes) fully connected by edges; self-loops are not shown here. Two nodes ($i$ and $j$) and the edge connecting them ($ij$) are highlighted. (b) $\mathbb{E}(3)$-invariant edge features are constructed by selecting an (arbitrary) origin and a pair of nodes, and computing the distance $d_{ij}$, inner product between unit vectors $u_i$ and $u_j$, and inner product between unit vectors $u_i$ and $u_{i-j}$. (c) We show a flow diagram for the GNN, which receives inputs on the left and predicts outputs on the right. The graph interactions are composed of $\bm\phi$ and $\bm\psi$, learnable functions parameterized by neural networks, and $\bm\oplus$, a permutation-invariant function that aggregates edge information onto a single node.
    We use multiple ``unshared'' edge layers that compute features in parallel.
    See text for more details.
}
    \label{fig:gnn-schematic}
\end{figure*}

A graph is a mathematical structure comprising a set of objects and the relationships between those objects.
For each TNG300 sub-box, we construct a cosmic graph to represent all subhalos and their pair-wise relationships. 
Subhalos are represented as nodes, and pairs of subhalos are connected with edges if they are separated by less than $L$, the linking length.
We do not permit nodes to connect to themselves (i.e., no self-loops).
The ordering of subhalos is irrelevant, as graphs are invariant (and their nodes are equivariant) to permutations of the node indices.
We will typically denote nodes or node properties using a single index, such as $i$, and edges or edge properties with two indices, such as $ij$. 
An example graph is shown in Figure~\ref{fig:gnn-schematic} panel~(a).

Statistical and ML models can harness the symmetries and inductive biases of mathematical graphs and efficiently learn robust representations \citep[e.g.,][]{Battaglia+2018}.
One important characteristic of graphs and sets is permutation invariance: the data have no natural ordering, so shuffling the node indices has no effect.
Additionally, subhalos reside in a $3d$ space and obey geometric constraints, i.e., they are invariant under the $\mathbb{E}(3)$ group action.
GNNs can learn robust representations from fewer data examples by imposing the graph structure and various symmetries as \textit{constraints} on the model \citep{Villar+2021,GeigerSmidt2022}.
These symmetries are due to physically invariant or equivariant phenomena, and models that have such symmetries inherent to them can learn more efficiently. 
In Section~\ref{ssec:inductive-biases}, we discuss the inductive biases of GNNs in more detail.
We also refer the interested reader to \citet{Battaglia+2018}, \citet{Schlichtkrull+2018}, and \citet{Bronstein+2021}, which describe GNNs and geometric deep learning at a high level.

\subsubsection{Graph features} \label{sssec:gnn-features}

We can ascribe features to graph nodes and edges.
For nodes, we only use the subhalo mass, maximum circular velocity, and whether the subhalo is a central or satellite.
Importantly, we do not assign positions or velocities as node features since they depend on the frame of reference.
Instead, we construct edge features that are fully invariant under the $\mathbb{E}(3)$ group action: three features based on pairs of node positions and three features based on pairs of node velocities.
Our choice of architecture and feature set ensures that our model is an $\mathbb{E}(3)$-invariant GNN \citep[for more, see e.g.,][]{GarciaSatorras+2021,Villar+2021}, and differs from the model used by \citet[][]{WuJespersen2023} because we include invariant features in velocity space.

To create these edge features, we follow \cite{Villanueva-DomingoVillaescusa-Navarro2022}, and first define unit vectors $\bm{u_i} \equiv (\bm{x_i} - \bm{\bar{x}}) / ||\bm{x_i} - \bm{\bar{x}}||$ relative to the center of the distribution $\bm{\bar{x}}$ (which we treat as an arbitrary origin).
We then create the following features between nodes $i$ and $j$: the squared Euclidean distance $d_{ij} \equiv || x_i - x_j ||$, the inner product $\alpha_{ij} \equiv \bm{u_i} \cdot \bm{u_j}$, and the inner product $\beta_{ij} \equiv \bm{u_i} \cdot \bm{u_{i-j}}$.\footnote{$\bm{u_{i-j}}$ is the unit vector toward $\bm{x_i}-\bm{x_j}$.}
If a triangle is drawn using the origin and nodes $i$ and $j$, then $\alpha_{ij}$ and $\beta_{ij}$ can be geometrically interpreted as the cosine of the angles opposite and next to the edge $ij$.
Figure~\ref{fig:gnn-schematic} panel~(b) illustrates these edge features.

\subsubsection{GNN architecture} \label{sssec:gnn-architecture}

As discussed in the previous section, we first create our cosmic graph with node features 
\begin{equation}    
X_i = \left (M_{{\rm halo}, i}, V_{{\rm max}, i}, \Gamma_{{\rm cen}, i} \right ),
\end{equation}
and edge features 
\begin{equation}
\mathcal E_{ij} = \left (d_{ij}^{(\bm{x})}, \alpha_{ij}^{(\bm{x})}, \beta_{ij}^{(\bm{x})}, d_{ij}^{(\bm{v})}, \alpha_{ij}^{(\bm{v})} \beta_{ij}^{(\bm{v})} \right),
\end{equation}
between any two nodes that are separated by less than the linking length, $L$.
These features are passed through our GNN, which is trained to predict 
\begin{equation}
y_i = \left (M_\star, \sigma_{M_{\star}} \right ), 
\end{equation}
i.e., the stellar mass and the scatter on stellar mass for every subhalo.
A schematic of our GNN architecture is shown in Figure~\ref{fig:gnn-schematic} panel~(c), while details are presented in Appendix~\ref{app:GNN-details}.

Our GNN is parameterized by learnable functions $\bm\psi$ and $\bm\phi$, which are multilayer perceptrons (MLPs).
We also use $N_{\rm unshared}=8$ MLPs in parallel to learn interactions on graph edges, and these are known as ``unshared'' layers.
These unshared layers are indexed by $\ell = \{1, 2, \cdots, N_{\rm unshared}\}$.

For some edge indexed by $ij$, we provide node features $X_i$, $X_j$, and edge features $\mathcal{E}_{ij}$ as inputs to a function, $\bm\phi$, which produces a hidden state $e_{ij}^{(\ell)}$ per unshared layer:
\begin{equation}
    e_{ij}^{(\ell)} = \bm\phi^{(\ell)}(X_i, X_j, \mathcal E_{ij}).
\end{equation}
The MLP $\bm\phi^{(\ell)}$ comprises two linear layers, each with $N_{\rm hidden}=64$ hidden neurons, LayerNorm, and then SiLU activation, and outputs a latent vector with $N_{\rm latent}=16$ dimensions via a final linear layer.

An aggregation function $\bm\oplus_j$ operates on all edge hidden states $e_{ij}$ that connect to node $i$, i.e., it pools over all neighboring $j$.
Here we define $\bm\oplus$ to be a multi-pooling operator, which concatenates the outputs of sum-pooling, mean-pooling, and max-pooling functions.
Since these pooling functions are invariant to permutations, the multi-pooling operator is also permutation invariant.

The function $\bm\psi$ receives a concatenated set of $N_{\rm unshared}$ features pooled into node $i$, as well as the node variable $X_i$, to make predictions:
\begin{equation}
    y_i = \bm\psi \left (X_i, \left\{\bm\oplus_j \left (e_{ij}^{(\ell)} \right ) \right\}_{(\ell)} \right).
\end{equation}
The function $\bm\psi$ is composed of three parts: $\bm \psi_1$, which ingests all pooled edge features and returns a 16-dimensional latent node state, $\bm \psi_2$, which ingests the latent node state and outputs a 16-dimensional latent node state, and $\bm \psi_3$, which ingests the concatenated outputs of $\bm \psi_1$ and $\bm \psi_2$ and produces the final prediction.
Each of $\bm \psi_1$, $\bm \psi_2$, $\bm \psi_3$ have two hidden layers, each with $N_{\rm hidden}=64$ hidden neurons, LayerNorm, and SiLU activations, and a final linear layer; they only differ in the number of input and output neurons.

We note that a deeper neural network can be constructed by sequentially stacking GNN layers. 
For example, if we instead predicted another node latent state and allowed a second round of interactions using learned functions across the graph of node and edge latent states, then we could represent higher-order interactions up to a length scale $2 L$. Increased GNN depth would enable improved performance due to enhanced internal representations of neighboring galaxies before the final layer.
Indeed, several works find that increasing the number of sequential GNN layers can improve the predictive power \citep[e.g.,][]{Sanchez-Gonzalez+2020,Pfaff+2021,Lam+2022}.
However, we are interested in assessing the optimal linking length scale, so we only use a single GNN layer in depth, thereby restricting the receptive field to a radius of $L$.

\subsubsection{GNN optimization procedure}
The data are split into training and validation sets using $k=3$-fold cross-validation as follows.
First, we divide the TNG300 box along the $z$ dimension into three equal sub-volumes.
We train on two sub-volumes and validate on the remaining sub-volume.
The TNG300 box is periodic, so the two training sub-volumes are always contiguous.
During training, we remove $10$~Mpc from the box as ``padding'' along the $z$ axis, which ensures that the training and validation sets are independent.
In other words, the training sub-volume is $\sim 300 \times 300 \times 180$ Mpc$^3$, while the validation sub-volume is $\sim 300 \times 300 \times 100$ Mpc$^3$.
We iterate over the three validation sub-volumes and concatenate the validation results to catalog predictions over the entire volume without any gaps.
Results are reported as the mean and standard deviation of the three cross-validation predictions unless otherwise noted.

We divide the data into mini-batches using the Pytorch-Geometric  \texttt{ClusterLoader} sampling strategy \citep{Chiang+2019} with 48 clusters.
At each mini-batch optimization step, GNN model parameters are updated to minimize a negative log-likelihood loss:
\begin{equation}
    {\rm NLL} = \frac{1}{2b} \sum_{i=1}^{b} \frac{(M_{\star, i} - \hat M_{\star, i})^2}{\hat \sigma^2_{M_\star}} -  \frac{1}{2} \log \left ( \hat \sigma^2_{M_\star}\right ) ,
\end{equation}
where $b$ is the batch size and $\hat \sigma^2_{M_\star}$ is the the predicted variance averaged over a batch.
We train the GNN for 300 epochs, where an epoch represents a full pass through the training set, and then report the predicted $M_\star$ and $\sigma_{M_{\star}}$ for the validation set.
We describe additional GNN architecture and training procedure details in Appendix~\ref{app:GNN-details}.

\section{Results: Model Performance} \label{sec:results}

\begin{deluxetable}{lrrc}
    \tablehead{
    \colhead{Model} & $N_{\rm features}$ & \colhead{$L_{\rm best}$} & \colhead{RMSE} \\
     & & (Mpc) & (dex) 
    }
    \startdata
    SHAM-$V_{\rm max}$ & 1 & --- & 0.1779 $\pm$ 0.0007 \\
    \EBMbase{}        & 3 & --- & 0.1660 $\pm$ 0.0011 \\
    \EBMdisperse{}    & 8 & --- & 0.1618 $\pm$ 0.0013 \\
    \EBMoverdensity{} & 4 & \DLinkOverdensity{} & 0.1606 $\pm$ 0.0012 \\
    \textbf{GNN}              & \textbf{12} & $\bm{\DLinkGNN{}}$ & ~$\bm{0.1499 \pm 0.0017}$
\tablecaption{\textbf{Comparison of main ML model results.} We show how different models perform on estimating $M_\star$ from subhalo properties. 
We have selected the best environmental length scale for the GNN and EBM with spherically averaged overdensity ($\delta_L$) models. 
The RMSE shown is the mean and standard deviation using $k=3$-fold cross-validation. The best performance is highlighted in bold.
\label{tab:comparison-results}
}
\enddata
\end{deluxetable}

We train the \EBMbase{}, \EBMdisperse{}, \EBMoverdensity{}, and GNN models to convergence, as described in the previous section.
The \EBMoverdensity{} and GNN experiments are repeated for multiple values of $L$, the environmental length scale.
We compare the stellar mass root mean squared error (RMSE),
\begin{equation}
    {\rm RMSE} = \left (\frac{1}{N} \sum_{i=1}^{N} \left (M_{\star, i} - \hat M_{\star, i} \right )^2 \right)^{1/2},
\end{equation}
as the RMSE directly quantifies the prediction accuracy.

We summarize each of the best model's results in Table~\ref{tab:comparison-results}.
For each model, we show the number of input features ($N_{\rm features}$) and the optimal linking length ($L_{\rm best}$).
We also show results for a SHAM model that matches $V_{\rm max}$ to $M_{\star}$, which serves as a commonly used baseline model \citep[SHAM-$V_{\rm max}$, e.g.,][]{Conroy+06}.
Although the EBM models outperform the SHAM-$V_{\rm max}$ model, the EBM performance can be further stratified: the \EBMdisperse{} and \EBMoverdensity{} models have similar performance and surpass the \EBMbase{} model.
The GNN performance handily exceeds all others.

\begin{figure}
    \centering
    \includegraphics[width=\columnwidth]{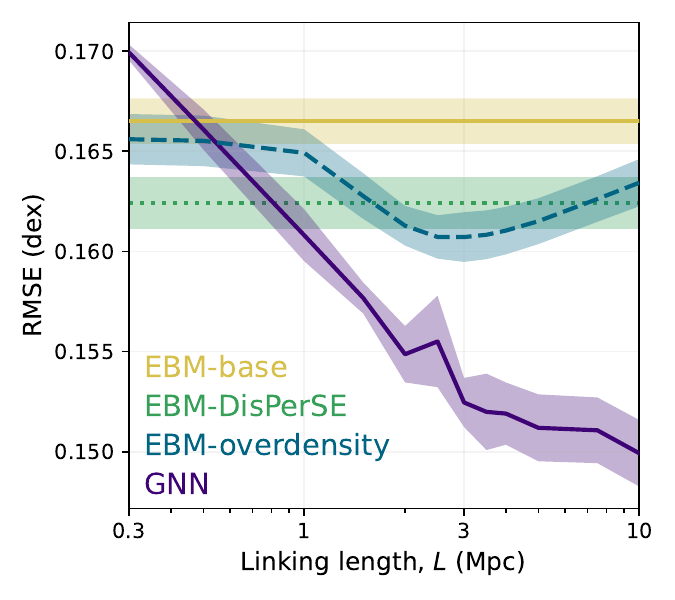}
    \caption{\textbf{Stellar mass prediction errors (lower is better) for different models as a function of $\bm{L}$. GNNs with large linking lengths perform the best.} We show model performance for EBM-base (solid yellow), EBM-DisPerSE (dotted green), EBM-overdensity (dashed blue), and GNN (solid purple). Lines and shaded regions depict the mean and $1\sigma$ scatter on the RMSE computed from $k=3$ cross-validation}.
    \label{fig:optimal-linking-length}
\end{figure}

Figure~\ref{fig:optimal-linking-length} shows how stellar mass estimates depend on linking length.
The \EBMbase{} and \EBMdisperse{} models do not vary with linking length.
For the \EBMoverdensity{}, the error is minimized at a linking length of \DLinkOverdensity{}~Mpc.
At smaller or larger scales, the average spherical overdensity is less useful for predicting stellar mass.

For the GNN model, we find that the RMSE decreases with increasing linking length until $\sim 3$~Mpc, and then modestly  at $L > 3$~Mpc (with low statistical significance). 
Whereas the \EBMoverdensity{} smooths out structure at $L > 3$~Mpc and therefore \textit{loses} information on smaller scales, the GNN model can continue to model environmental impacts at all scales up to $L$.

\section{EBM Model Interpretation} \label{sec:interpretation}

A critical part of any ML analysis is to interpret and diagnose the models \citep{Huppenkothen+2023}.
While there exist methods that attempt to explain the decision-making processes of ``black-box'' ML algorithms such as neural networks \citep[e.g.,][]{saliency}, many of these explanations are subjective and require yet another layer of interpretation. 
Instead, inherently interpretable models should be used when possible \citep[see, e.g.,][]{Rudin2019}.
In this section, we explore physical interpretations of our EBM model results.

\subsection{Feature importance with SHAP} \label{ssec:shap}

EBMs are interpretable because they are additive models (see Section~\ref{ssec:ebm}).
One can directly plot each feature's contribution to the final prediction, although the interpretation can still be subjective. 
Using terminology from Equation~\ref{eq:GAAM}, we could plot $f_i(x_i)$ versus $x_i$ and $f_{ij}(x_i, x_j)$, marginalized over $x_j$, versus $x_i$ to see how the model prediction depends on $x_i$, i.e., visualization can be used for \textit{qualitative} assessments.
However, here we would like to \textit{quantitatively} interpret the importance of each feature. 

To measure feature importances, we use a method called SHAP, or SHapley Additive exPlanations \citep{SHAP}.
SHAP computes each feature's contribution to the prediction based on cooperative game theory \citep[building on another metric called Shapley values; see][]{shapleyvalues}. 
SHAP returns the fair distribution of additive contributions even when the model contains correlated features. 
Exact Shapley values can be computed using permutations of all features, which is computationally infeasible; we instead follow the standard method of randomly sampling examples with replacement ($N=$~5,000).
We use the samples to approximate Shapley values and compute their SHAP feature importances. 
For a more technical overview of SHAP and Shapley values, we refer the interested reader to Sections 9.5 and 9.6 of \cite{molnar2022}, and the original SHAP paper \citep{SHAP}, which describes several additional desiderata.
We implement the algorithm using the \texttt{shap} Python package.

\subsection{Are cosmic web features more informative than overdensity?} \label{ssec:ebm-all}

\begin{figure}
    \centering
    \includegraphics[width=\columnwidth]{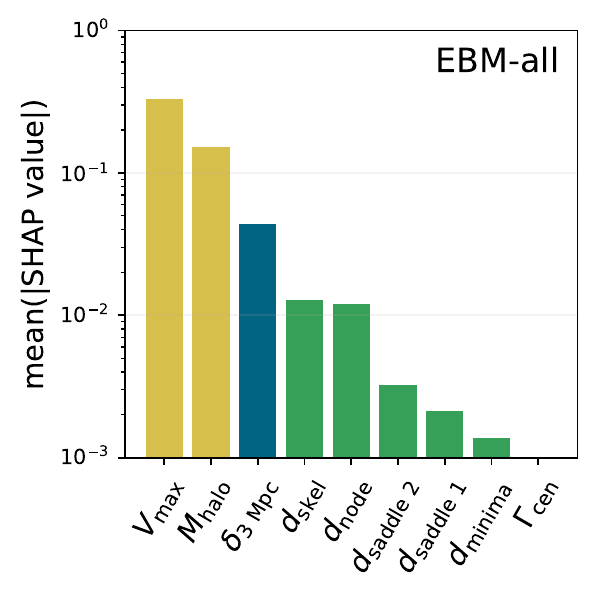}
    \caption{
    \textbf{After \bm{$V_{\rm max}$} and \bm{$M_{\rm halo}$}, the most important EBM feature is spherical overdensity.} We show SHAP values for \EBMall{}, which predicts stellar mass using an EBM model with base halo features (yellow), DisPerSE cosmic web features (green), and overdensity on 3~Mpc scales (blue).
    A higher mean absolute SHAP value indicates greater feature importance.
    \label{fig:ebm-shap}
    }
\end{figure}

We cross-validate a new model, \EBMall{}, which includes the set of all features in \EBMdisperse{} and \EBMoverdensity{} at $L=3$~Mpc. 
In other words, the model is given the base subhalo properties, DisPerSE features, and the $\delta_{3~\rm Mpc}$ averaged overdensity feature. 
This \EBMall{} model achieves RMSE = $0.1596 \pm 0.0012$, which is on par with the other top-performing EBM models but significantly less accurate than the GNN.
Crucially, we can use SHAP to compare the relative importance of different model features.

We define feature importance as the mean absolute SHAP value, i.e., mean($\vert$SHAP value$\vert$).
We use the mean \textit{absolute} values because the features can add to or subtract from the final prediction, and we do not wish to average out their positive and negative contributions.

In Figure~\ref{fig:ebm-shap}, we present a bar chart of the most important \EBMall{} features.
We find that $V_{\rm max}$ and $M_{\rm halo}$ are most important for estimating $M_\star$, followed by overdensity $\delta_{3~\rm Mpc}$, and then the distance to the cosmic web skeleton and nodes.
Remarkably, the 3~Mpc scale overdensity is more important than the summed feature importances of \textit{all} cosmic web distances (note that Figure~\ref{fig:ebm-shap} shows feature importance on a logarithmic scale).
This result is consistent with our earlier finding that the \EBMoverdensity{} model at 3~Mpc outperforms the \EBMdisperse{} model (at low significance; see Figure~\ref{fig:optimal-linking-length}).
Although the DisPerSE hyperparameters are not tuned for our stellar mass prediction task, it is surprising nonetheless that the simple spherical overdensity feature outperforms the sophisticated cosmic web distance features.
Our experiments show that overdensity is more informative than the DisPerSE features for predicting stellar mass from TNG300 subhalo catalogs.

Intriguingly, we find a very low SHAP value corresponding to whether a galaxy is in a central or satellite halo, $\Gamma_{\rm cen}$. 
Although we would expect this variable to improve the model's predictive power, we have also seen that separate central and satellite galaxy--halo relationships can be learned without explicitly accounting for $\Gamma_{\rm cen}$ (e.g. \citealt{WuJespersen2023}; see also \citealt{Behroozi2019}).
Additionally, $\Gamma_{\rm cen}$ may have significant redundancies or interactions with other features, which could lead to a depressed SHAP value.

\subsection{What is the most important overdensity scale?} \label{ssec:interpret-length-scale}

\begin{figure*}[ht]
    \centering
    \includegraphics[width=0.495\textwidth]{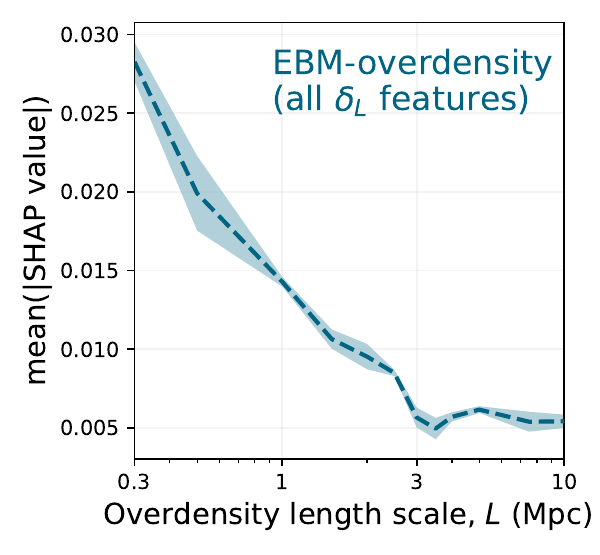}
    \includegraphics[width=0.495\textwidth]{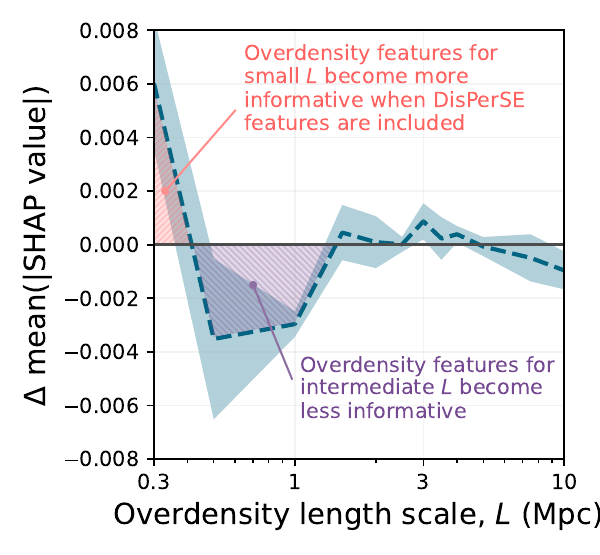}
    \caption{\textbf{The most important length scales for overdensity (left), and changes to those relative importances when DisPerSE features are added (right).} In the \textit{left} panel, we show SHAP values for an EBM model that employs multiple overdensity features, each with different length scales $L$. SHAP values for $M_{\rm halo}$ and $V_{\rm max}$ are not shown here. %
    In the \textit{right} panel, we include DisPerSE parameters in addition to all overdensity and base features in the model fit and interpretation, which increases the relative importance of $\delta_L$ for small length scales, and decreases the relative importance of $\delta_L$ for intermediate length scales.
    }
    \label{fig:ebm-Dlink-shap}
\end{figure*}

We cross-validate a new model like the \EBMoverdensity{} model, except that we include overdensity for all $L \in \{0.3, 0.5, 1, \ldots, 10\}$~Mpc as separate features.
This model with all  $\{\delta_{L}\}$ and base subhalo features ($M_{\rm halo}, V_{\rm max}, \Gamma_{\rm cen}$) achieves a best RMSE of $0.1592 \pm 0.0017$.
Its performance is comparable or slightly better than the other EBM models, but still significantly worse than the GNN.
Again, we can compare the feature importances for different $\delta_L$.

In the left panel of Figure~\ref{fig:ebm-Dlink-shap}, we show the feature importance of $\delta_L$ for the \EBMoverdensity{} model varied over different values of $L$ (using $k=3$-fold cross-validation as before).
We find that the overdensities on smaller scales are more informative in terms of SHAP values.
As $\delta_L$ extends toward larger scales, the feature importance diminishes until it hits a plateau at $L \approx 3$~Mpc.
Beyond $L > 3$~Mpc, the feature importance stays at a relatively low constant level.
Nonetheless, SHAP indicates that there is useful information at large scales, even at $10$~Mpc.

Including overdensity over progressively larger scales yields diminishing returns as we smooth out information.
This trend of declining feature importance with overdensity length scale is remarkably similar to the results shown in Figure~\ref{fig:optimal-linking-length}.
Just as the GNN performance improvements diminish and the \EBMoverdensity{} model performance suffers above $L > 3$~Mpc, Figure~\ref{fig:ebm-Dlink-shap} shows that the feature contribution of $\delta_L$ also plateaus above $L>3$~Mpc length scales.
Therefore, we interpret these findings as evidence that the most important environmental overdensity in TNG300 is on $3$~Mpc length scales.

To understand the interplay between spherical overdensity and cosmic web environmental parameters, we repeat the previous experiment while including DisPerSE parameters.
The cross-validation RMSE is the same as before, $0.1592 \pm 0.0017$, which implies that SHAP feature importances can be directly compared between the two models.
Because including DisPerSE features has no impact on the cross-validation loss, the overdensity and cosmic web features are somewhat redundant.
We plot the difference in \textit{relative} feature importance for overdensity, $\Delta$ mean($\vert$SHAP value$\vert$), for overdensity against $L$ in the right panel of Figure~\ref{fig:ebm-Dlink-shap}.
We find a significant increase in relative feature importance for overdensity at $\sim$0.3~Mpc scales and a decrease in relative feature importance for overdensity at 0.5~Mpc~$\lesssim L \lesssim$~1.5~Mpc scales.
These are illustrated using the figure's red and purple shaded regions, respectively.
We conclude that DisPerSE cosmic web features primarily capture information on intermediate scales ($\sim 1$~Mpc) rather than very large scales ($\gtrsim 3$~Mpc).

\section{Discussion} \label{sec:discussion}

Our EBM and GNN model results show that large-scale environmental features improve models of the TNG galaxy--halo connection by a significant margin.
The \EBMbase{} model can be augmented with cosmic web features or an averaged spherical overdensity, or better yet, a neural network can flexibly learn environmental parameters by representing subhalos as a cosmic graph.
While these advanced methods are capable of lowering the RMSE for stellar mass estimates (Section~\ref{sec:results}) and for interpreting the model results (Section~\ref{sec:interpretation}), there are many other important ways to evaluate these methods' scientific utility.
This section discusses other summary statistics and considerations in cosmology and astrophysics.

\subsection{Stellar mass function}

The stellar mass function---the distribution function of galaxies by stellar mass---is one of the most essential metrics for comparing galaxy populations \citep[e.g.,][]{Weaver2023_COSMOS_GSMF}.
The stellar mass function depends on physical processes ranging over a variety of scales, from cloud-scale gas cooling and star formation to large-scale galaxy interactions and tidal torques.
ML methods that aim to paint galaxy properties onto subhalos should be able to reconstruct the stellar mass (or luminosity) function accurately.

\begin{figure}
    \includegraphics[width=\columnwidth]{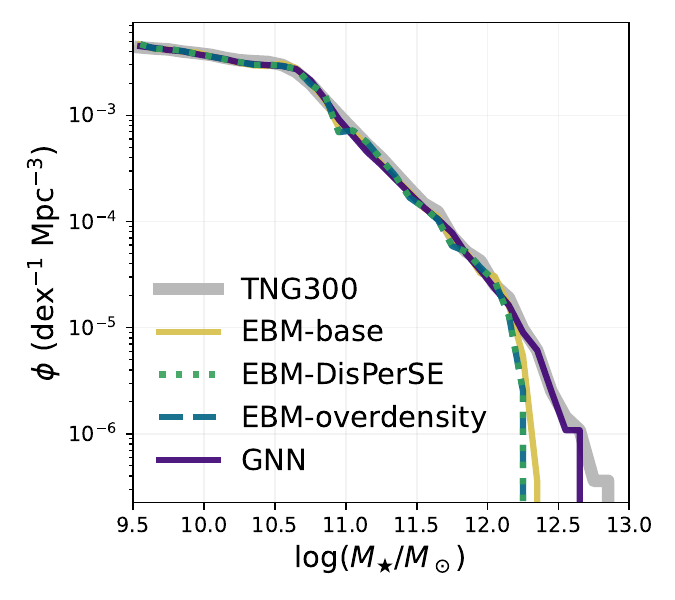}
    \caption{\textbf{GNNs surpass EBMs in recovering the stellar mass function.} We show stellar mass functions, $\phi$, computed using the EBM and GNN methods and compared against the TNG300 simulation ground truth. }
    \label{fig:stellar-mass-function}
\end{figure}

In Figure~\ref{fig:stellar-mass-function}, we show the stellar mass function, $\phi$, from \EBMbase{} (solid yellow), \EBMdisperse{} (dotted green), \EBMoverdensity{} (dashed blue), and GNN (solid purple) predictions.
We show the TNG300 ground truth stellar mass function in thick solid gray.
All the ML models predict accurate stellar mass functions in the range $9.5 < \log(M_\star/M_\odot) < 12$, demonstrating their general success at estimating galaxy stellar mass at the population level. 
While the EBM models begin to falter above $\log(M_\star/M_\odot) \approx 12.25$, the GNN continues to recover the stellar mass function up to $\log(M_\star/M_\odot) \approx 12.6$.
These underpredictions reflect the well-known limitation of ML models to capture the extrema of the target distribution, where the training data offer very few examples.
Indeed, the most massive galaxies are only found at the centers of rare, massive galaxy clusters \citep[and tend to have very extended and diffuse stellar mass distributions; e.g.,][making their properties difficult to model]{Pillepich+2018}.
For these extreme examples, it is often expected that any statistical model performance will suffer or break down (signifying a physical shift in algorithmically learned scaling relations; e.g., \citealt{Wu2020}).\footnote{In fact, if ML models do \textit{not} break down at the extrema, then this could indicate a surprising continuation of scaling laws into extremal regimes even when new physical processes might become important \citep[e.g.,][]{Holwerda+2021}.}
However, we confirm that the EBM and GNN models can accurately reproduce the stellar mass function for common, well-represented galaxies.

\subsection{Two-point correlation function} 

\begin{figure*}
    \centering\includegraphics[width=0.495\textwidth]{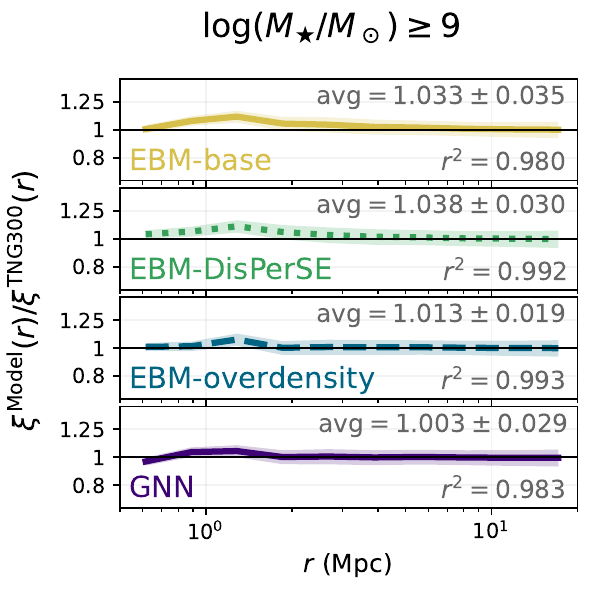}
    \includegraphics[width=0.495\textwidth]{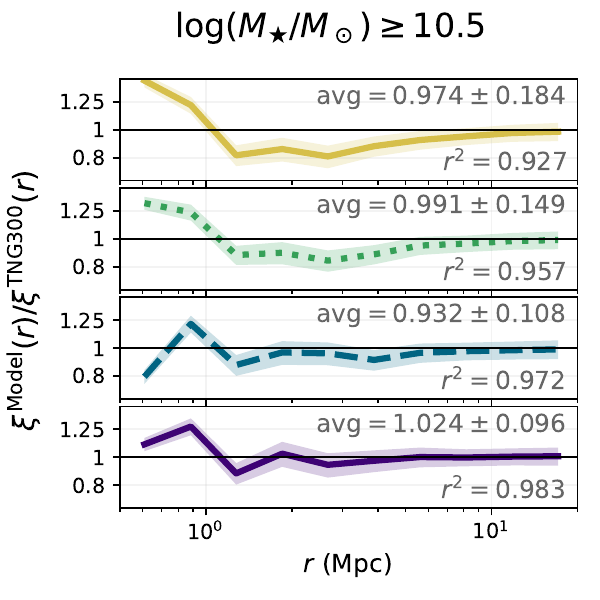}
    \caption{
    \textbf{Comparing different models' two-point correlation functions.} The real space galaxy-galaxy correlation function $\xi^{\rm Model}(r)$ from EBM and GNN models, normalized by the ``true'' TNG300 correlation function $\xi^{\rm TNG300}(r)$, for low-mass (\textit{left}) and high-mass (\textit{right}) bins. The shaded regions show uncertainties from shot noise and marked bootstrap resampling. We also display the weighted average $\xi^{\rm Model}(r)/\xi^{\rm TNG300}(r)$ and the coefficient of determination, $r^2$.
    }
    \label{fig:2pt-corr-ratio}
\end{figure*}

Methods that use (small-scale) environment information can affect downstream clustering analyses, which in turn can introduce biases in cosmological applications.
We test whether our EBM or GNN approach corrupts the galaxy clustering information by measuring real space galaxy-galaxy correlation functions, $\xi(r)$, using the \texttt{treecorr} code.
The count-count correlations are measured in logarithmically spaced radial bins from $0.5$ and $20$~Mpc.\footnote{At scales $>20$~Mpc, all models can almost perfectly recover the TNG300 two-point correlation function.}
We split galaxies into two stellar mass samples, $9 \leq \log(M_\star/M_\odot)<10.5$ and $\log(M_\star/M_\odot) \geq 10.5$, where $M_\star$ is independently predicted using each ML model, optimized using the full training sample.
We normalize two-point correlation functions by the ground truth two-point correlation function, $\xi^{\rm TNG300}(r)$, based on the TNG300 fully hydrodynamic simulation.
We estimate uncertainties using shot noise variance and the covariance derived from marked bootstrap resampling \citep[see][]{Loh2008}.
We also report two metrics: first, we compute the weighted mean ratio of $\xi^{\rm Model}(r)/\xi^{\rm TNG300}(r)$, which measures the level of bias in modeled correlation function. 
Second, we compute the coefficient of determination, $r^2$, between each $\xi^{\rm Model}(r)$ and $\xi^{\rm TNG300}(r)$, which measures how well the two correlation functions agree.
Our results are shown in Figure~\ref{fig:2pt-corr-ratio}.

We find that all ML models generally reproduce the correct two-point correlation function, especially for large separations ($> 10$~Mpc), which should not be surprising given that these scales are in the linear regime.
The models that leverage environmental features perform as well or better than the \EBMbase{} model.
For separations smaller than a few Mpc, and for higher-mass galaxies, the errors on model-predicted correlation functions range up to $\sim 0.1$~dex (panel~(b) of Figure~\ref{fig:2pt-corr-ratio}).
At $r \gtrsim 1$~Mpc, the \EBMbase{} and \EBMdisperse{} models cluster lower-mass galaxies too strongly, and higher-mass galaxies too weakly.
The \EBMoverdensity{} and GNN models are less prone to these small biases and give robust results.
The GNN appears to perform the best in terms of the mean ratio of correlation functions and $\rho$, although there is still some scatter in the higher-mass bin.
Ultimately, the two-point correlation function does not appear to be highly sensitive to differences between our model predictions.
We conclude that our methods do not introduce strong biases in the correlation statistics and that a GNN overall performs best.

\subsection{SHMR dependence on overdensity}

\begin{figure*}[ht]
    \centering
    \includegraphics[width=\textwidth]{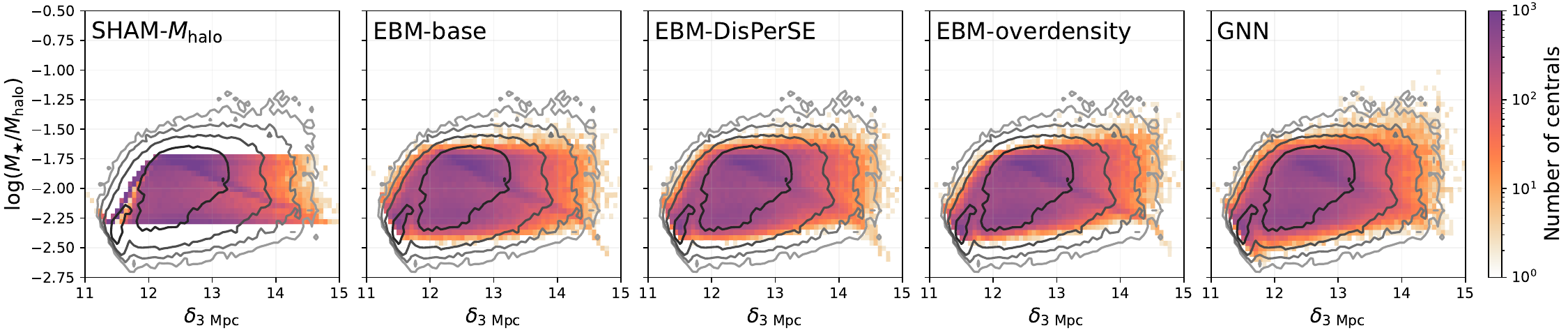}
    \includegraphics[width=\textwidth]{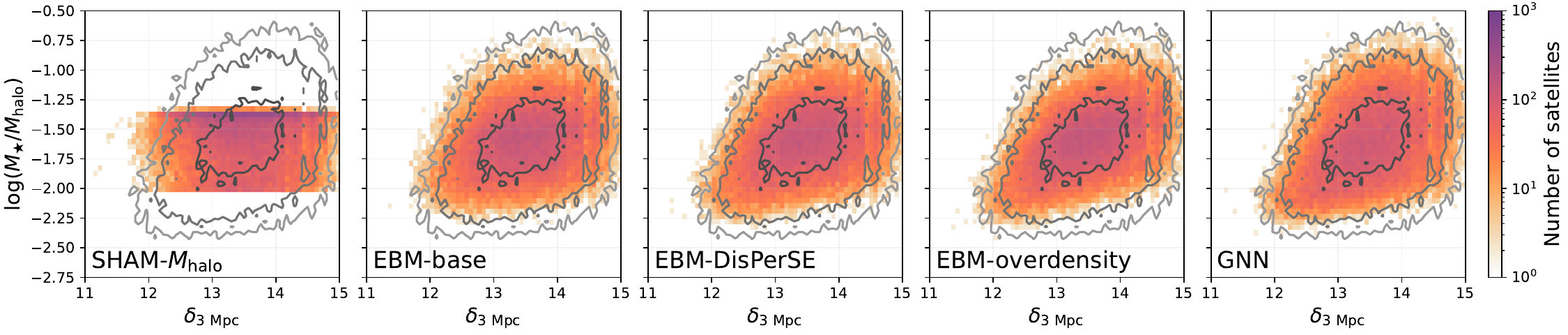}
    \caption{
    \textbf{In a comparison of $\bm{M_\star/M_{\rm halo}}$ against overdensity, the GNN is most similar to TNG300 (contours).} We show the stellar-to-halo mass ratio against average overdensity on 3~Mpc scales, for centrals (\textit{upper}) and satellites (\textit{lower}). The gray contours show the same comparison for TNG300 with levels at $4, 16, 64, 256$. All panels are shown on the same scales.
    }
    \label{fig:shmr-overdensity}
\end{figure*}

We also investigate how the modeled SHMR varies with environment.
Our results and interpretation in Section~\ref{ssec:interpret-length-scale} showed that large-scale surroundings on 3~Mpc scales are particularly influential for the SHMR, so we parameterize the environment using $\delta_{\rm 3~Mpc}$.
In Figure~\ref{fig:shmr-overdensity}, we compare the stellar-to-halo mass ratio, $\log(M_\star / M_{\rm halo})$ against $\delta_{3~\rm Mpc}$ for both central (\textit{upper}) and satellite (\textit{lower}) subhalos.
We plot this relationship for a simple $M_{\rm halo}$ SHAM model (\textit{left}-most panel), EBM models (\textit{middle} panels), and the GNN (\textit{right}-most panel).
The ground truth TNG300 simulation is shown in all panels as gray contours.

We see a strong ``ridge line'' in $\log(M_\star / M_{\rm halo})$ versus $\delta_{3~\rm Mpc}$ in central halos for some models (\textit{upper} panels in Figure~\ref{fig:shmr-overdensity}).
The ridge is inhabited by lonely central halos with no other subhalos within 3~Mpc, such that the overdensity is equal to the central halo mass.
The monotonic SHMR from abundance matching leads to a particularly tight ridge line (left-most panel).
The GNN model is vastly better than EBM models at softening this ridge line.
However, we note that the ridge is not a complete artifact, and manifests as a broad trend even in the TNG300 version of the $\log(M_\star / M_{\rm halo})$ versus $\delta_{3~\rm Mpc}$ relationship for centrals.
We conclude that the GNN most realistically recovers this probe of the galaxy--halo--environment connection.

\subsection{Variations in the optimal GNN linking length} \label{ssec:linking-length-variations}

\begin{figure*}[ht]
    \centering
    \includegraphics[width=\textwidth]{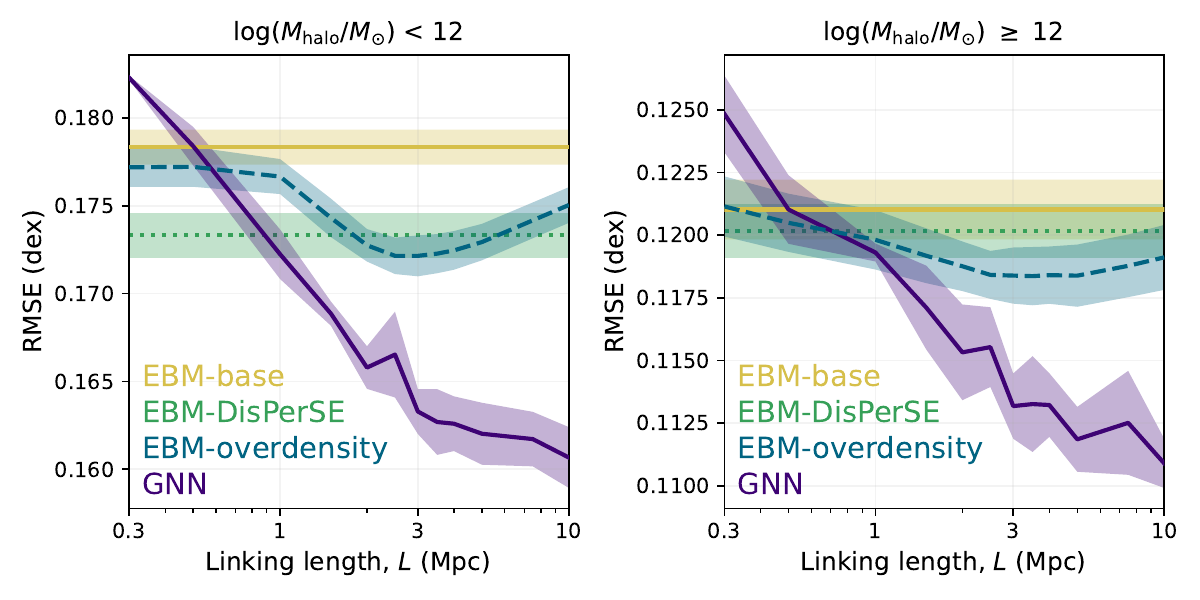}
    \caption{
    \textbf{The \EBMoverdensity{} and GNN models demonstrate improved performance at larger \bm{$L$} for higher-mass subhalos.} We show how different models perform as a function of $L$ for lower-mass (\textit{left}) and higher-mass (\textit{right}) subhalos. Note that the $y$ axes show different scales.}
    \label{fig:split-by-halo-mass}
\end{figure*}

A halo's assembly history is correlated with its mass and depends on whether it is a central or satellite.
We expect this to manifest as systematic variations in the optimal environmental length scale.
To test this, we bin the model predictions by subhalo mass, i.e. $\log(M_{\rm halo}/M_\odot) < 12$ and $\geq 12$.
A halo mass of $10^{12}~M_\odot$ is near the characteristic mass that creates a break in the SHMR \citep[see, e.g.,][]{Moster+2010,Engler+2021}.
These results are shown in Figure~\ref{fig:split-by-halo-mass}.
Note that we do not retrain the models, but simply split the validation results into low- and high-mass subsamples.
We also observe that the typical errors associated with stellar mass predictions in the two mass bins are different, ranging from $\gtrsim 0.17$~dex in the lower-mass bin to $\gtrsim 0.12$~dex in the higher-mass bin.
The \EBMoverdensity{} model is also less sensitive to linking length in the higher-mass bin.
Although we find qualitatively similar results to before, the optimal values of $L$ vary modestly with subhalo mass.
For \EBMoverdensity{}, these are 3~Mpc for $\log(M_{\rm halo}/M_\odot) < 12$, and 3.5~Mpc for $\log(M_{\rm halo}/M_\odot) \geq 12$. 
The GNN exhibits more steeply declining errors at large $L$ for the higher-mass bin, indicating that it can better leverage large-scale environmental information surrounding more massive subhalos.

\begin{figure*}
    \centering
    \includegraphics[width=\textwidth]{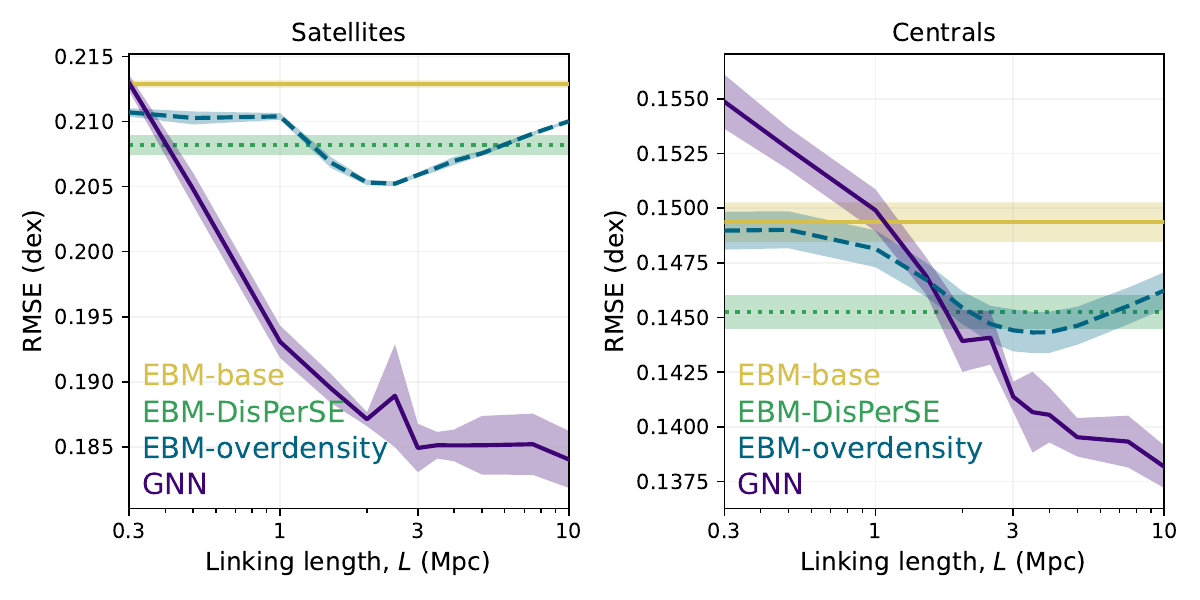}
    \caption{
    \textbf{The \EBMoverdensity{} and GNN models exhibit better performance at larger \bm{$L$} for centrals.} We show how different models perform as a function of $L$ for centrals (\textit{left}) and satellites (\textit{right}). Note that the $y$ axes show different scales.}
    \label{fig:split-by-central}
\end{figure*}

We also split the model prediction results by whether each subhalo is a satellite or central, shown respectively in left and right panels of Figure~\ref{fig:split-by-central}. 
We again find that the typical RMSE losses are different, and that they strongly vary with $L$.
Satellites appear to leverage information on smaller scales ($\sim 2.5$~Mpc) while centrals exploit information on larger scales ($\sim 3.5$~Mpc); this difference is particularly apparent from the \EBMoverdensity{} results.
We can also draw similar conclusions from the GNN results, which are largely flat for $L >3$~Mpc for satellites, in contrast to centrals which show continued improvement beyond $L > 3$~Mpc.
Our results demonstrate that the optimal $L$ systematically varies with a galaxy's halo mass and whether it is a central.

\subsection{Environmental distance scales in other works} \label{ssec:previous-works}

\cite{Lovell+2022} paint galaxies onto dark matter halos by training on the EAGLE simulations. 
Their approach is similar to our EBM models: by using extremely random trees and a larger set of halo features, they aim to estimate various galaxy properties (including stellar mass). 
They report that overdensity on $L=2-4$~Mpc scales are more important than overdensities on $1$ or $8$~Mpc scales; these findings are consistent with our results.
\cite{Wechsler2022} showed that a model that does not directly use halos but uses $\sim 5$ Mpc densities works well to describe the observed correlation function and color properties of observed galaxies.

Other works have investigated augmenting basic HOD models with environmental parameters in TNG simulations \citep[e.g.,][]{2019MNRAS.490.5693B,2020MNRAS.493.5506H,2021MNRAS.501.1603H,Delgado+2022,2022MNRAS.512.5793Y,2023MNRAS.524.2507H}. 
In general, $5-7$~Mpc scales are found to best describe overdensity.
HOD models are optimized to simultaneously predict the central and satellites, whereas we employ ML methods to predict a single central or satellite galaxy.
Thus, it is not surprising that the optimal environmental scale for HOD models is larger than our overdensity scale of $3$~Mpc.

Several studies have parameterized the large-scale environment to understand the correlation between large scale structure and other halo properties in purely $N$-body simulations.
\cite{2022MNRAS.516.5849R} find that tidal anisotropy on scales four times a halo's virial radius is optimal for describing large scale structure.
While this is a different task than modeling galaxies in halos, being able to predict properties such as halo spin and shape should have some bearing on predicting galaxy properties due to well-studied correlations \citep[e.g.,][]{2007MNRAS.378...55M}.

\subsection{Connection to galactic conformity?}

We have found that the connection between subhalos and galaxies depends critically on their surroundings on $\sim 3$~Mpc distance scales.\footnote{Nonetheless, the \EBMoverdensity{} and GNN models can exploit environmental information on scales larger than $3$~Mpc.}
This length scale is similar to galactic conformity correlation scales (beyond the central halo's virial radius, i.e., two-halo conformity).
One explanation for galactic conformity is that ``pre-heating'' from star formation or active galactic nuclei heats gas and inhibits future star formation, even if that gas is accreted into surrounding systems at later times \citep[e.g.,][]{Kauffmann+2013}.
\cite{Hearin+2016} posit that large-scale tidal forces are responsible for synchronizing halo accretion rates over several Mpc.
Other works find that gas reservoirs in both central and satellite galaxies can be stripped as they collectively enter through filaments or the outskirts of the most massive cluster-scale halos, which correlates their mass accretion and star formation histories \citep[e.g.,][]{Bahe+2013,Zinger+2018,Ayromlou+2023}.
Alternatively, \citet{ZuMandelbaum2018} contend that a simple halo quenching model, with no assembly bias or direct environmental processes, can fully reproduce the observed galactic conformity clustering signal \citep[see also, e.g.,][]{2018MNRAS.477..935T,2018ApJ...852...31W}.

Our study probes large-scale environment at a single simulation snapshot, and does not include any direct parameterization of assembly bias.
For example, we do not use peak halo mass or subhalo mass at time of accretion \citep[e.g.,][]{NagaiKravtsov2005,Conroy+06}, so our work only considers the galaxy--halo connection through the lens of large-scale environment.
Indeed, our results and interpretation in Section~\ref{sec:interpretation} demonstrate that environmental impacts can be captured without explicitly accounting for assembly bias \citep[although accounting for the halo merger history can further assist with stellar mass predictions; see, e.g.,][]{Jespersen+2022}.
However, we emphasize that these findings do not directly explain galactic conformity or its origin.

\subsection{The inductive biases of graph-based models} \label{ssec:inductive-biases}

Graphs are natural representations of cosmic structure, i.e., they capture the inductive biases of our problem.
There is no canonical ordering for a subhalo catalog, which is reflected in the permutation equivariance of graphs.
Interactions between galaxies and their surroundings are also local, which can also be enforced via graph connectivity on linking distance scales, $L$.
Additionally, graphs impose a relational structure between entities that are not reflected in other ML models, e.g., tabular models or fully connected neural networks \citep{2016MNRAS.457.1162K,2018MNRAS.478.3410A,2019MNRAS.490.2367C,Necib+2020,Wadekar+2020,Moster+2021,Delgado+2022,Bowden+2023,Hausen+2023}.
Our subhalo catalogs are actually point clouds in a $3d$ position space,\footnote{Point clouds have a notion of position, and therefore obey geometric inductive biases.} and we enforce these geometric constraints by making our graph models invariant under rotations, translations, and reflections (i.e., the $\mathbb{E}(3)$ group action).
Convolutional neural networks (CNNs) can also map baryons onto dark matter halos  \citep{2019arXiv190205965Z,Kasmanoff+2020,2022ApJ...941..132M}, and incorporate local relationships, but they make other assumptions about the geometric structure of the data, e.g., CNNs represent matter distributions on a coarse-grained grid.
Given the nature of our data, we strongly recommend using graphs or point clouds to model galaxy large-scale structure.

\subsection{Caveats} \label{ssec:caveats}

Here we mention several caveats about our methodology and analysis. 
First, we have only investigated the environmental length scales in TNG300, and not for other TNG box sizes or different simulation codes.
In particular, other simulations impose different physical models and subgrid prescriptions, which can cause differences in the galaxy--halo-environment connection.
Even for TNG300, we note that the optimal distance scale may change depending on the subhalo catalog selection criteria (see, e.g., Section~\ref{ssec:linking-length-variations}).
Different physics models or selection effects can lead to different ``optimal'' length scales.

In this work, we have only studied the $z=0$ TNG300 catalogs. 
It is likely that GNNs and other ML algorithms can learn salient environmental information from higher-$z$ subhalo catalogs, but we have not tested this regime. 
Moreover, it is possible that the optimal length scales for spherical overdensity ($3$~Mpc) and GNNs ($>10$~Mpc) may change at higher redshifts.

We also note that our use of DisPerSE distance features is limited to a single TNG300 value-added catalog \citep{Duckworth+2019}. 
This catalog is not meant to exhaustively cover all parameterizations of cosmic web features. 
It is also known that the DisPerSE algorithm can be tuned via several hyperparameters (such as the persistence threshold; \citealt{Sousbie2011}), and that these hyperparameter choices may impact the importance of \EBMdisperse{} model features.
Other studies have also run DisPerSE on TNG300 with different hyperparameter choices \citep[e.g., a different persistence threshold;][]{Galarraga-Espinosa+2020,Galarraga-Espinosa+2022}.

The final caveat stems from our subhalo catalog, which is crossmatched between the dark matter only and hydrodynamic simulations of TNG300 runs. 
We have used only the most robust subhalos found with \texttt{SUBFIND} and matched to dark matter only runs using both \texttt{LHaloTree} and \texttt{SubLink}, but the resulting subhalo population may still be biased or have incorrect properties \citep[see, e.g.,][]{Green+21,Benson+22,Mansfield+23}.
On top of this, there is a ``butterfly effect'' in numerical simulations that can contribute significant scatter to individual galaxy-to-halo mass ratios \citep{2019ApJ...871...21G}.
We do not attempt to correct for these issues, as they go beyond the scope of our aim to understand galaxy environmental length scales. 
See \citet{Chuang+23_IllustrisMangrove} for a further discussion of the impact of the butterfly effect on ML studies using simulations.
Our work is an initial foray toward understanding environmental length scales via explainable ML models and GNNs; additional studies are needed to confirm that these results generalize to other data sets.

\section{Conclusions} \label{sec:conclusions}

In this work, we investigate which 
distance scales and environmental measures best characterize the $z=0$ galaxy--halo--environment connection in the Illustris TNG300 simulation.
We learn to predict the stellar mass from dark matter halo properties using interpretable machine learning models (Explainable Boosting Machines; EBMs) and graph neural networks (GNNs; see Figure~\ref{fig:gnn-schematic} for a schematic diagram). 
Our analysis compares an \EBMbase{} model, an \EBMdisperse{} model with cosmic web distance features, an \EBMoverdensity{} model with spherically averaged overdensity features on length scale $L$ ranging between 0.3 through 10 Mpc, and a GNN with subhalos connected on those same length scales.
We evaluate each model using the root mean squared error (RSME) of stellar mass predictions compared to the hydrodynamic simulation, where the latter is treated as the ground truth.
Our code is publically available on Github at \url{https://github.com/jwuphysics/gnn-linking-lengths}.

Our main conclusions are as follows:
\begin{enumerate}
    \item An EBM model performs best if provided spherically averaged overdensity on 3~Mpc scales (Figure~\ref{fig:optimal-linking-length}).
    \item We interpret the EBM using SHapley Additive exPlanations (SHAP), and find that the overdensity features most important for describing the galaxy--halo connection are those on scales up to $\sim 3$~Mpc (Figure~\ref{fig:ebm-Dlink-shap}). At larger distances, overdensity is less informative but still contributes new information.
    \item If we train a model that includes the 3~Mpc overdensity feature as well as cosmic web features, then the simple overdensity feature is considerably more important than all cosmic web features (Figure~\ref{fig:ebm-shap}).
    \item Our GNN outperforms all other models by simply linking galaxies together on large scales (Figure~\ref{fig:optimal-linking-length}).
    \item The GNN performance continues to improve at $L \gtrsim 3$~Mpc, albeit at a slower pace, indicating that the GNN learns valuable large-scale features. These very large-scale environmental features are preferentially useful for modeling the galaxy--halo connection at high mass and for central halos (Figures~\ref{fig:split-by-halo-mass} and \ref{fig:split-by-central}, respectively).
\end{enumerate}

We ensure that our ML algorithms recover accurate stellar masses functions and two-point correlation functions (Figures~\ref{fig:stellar-mass-function} and \ref{fig:2pt-corr-ratio}, respectively). 
We also compare the stellar-to-halo mass ratio, $\log(M_\star/M_{\rm halo})$, against spherically averaged overdensity, $\delta_{\rm 3\ Mpc}$, which shows that the GNN most realistically accounts for the interplay between the SHMR and environmental overdensity (Figure~\ref{fig:shmr-overdensity}).

This work is an initial exploration of the environmental distance scales that influence the galaxy--halo relation in a specific model of galaxy formation and is not without its caveats and limitations (see, e.g., Section~\ref{ssec:caveats}).
However, we demonstrate that interpretable EBM models are capable of divulging important information about complex physical interactions.
An $\mathbb{E}(3)$-invariant GNN can outperform these EBM models simply by learning the galaxy--halo--environment connection from data; we do not need to explicitly provide the GNN any environmental features.
We believe that GNNs, with their relational inductive biases and group-invariant/equivariant constraints, will continue to lead the charge in efficient, interpretable ML for the physical sciences.
Follow-up studies with symbolic regression \citep[e.g., \texttt{PySR};][]{PySR} or other SHAP capabilities \citep[e.g., kernel explanations;][]{SHAP} can add another layer of interpretability to the work presented here. 
For understanding the features characterizing the galaxy--halo connection in the real universe, it will be important to expand this work beyond the single simulation considered herein to understand how generic these findings are for other models.

\software{
\texttt{astropy} \citep{astropy},
\texttt{InterpretML} \citep{InterpretML},
\texttt{matplotlib} \citep{matplotlib},
\texttt{gnn-linking-lengths} \citep{gnn-linking-lengths},
\texttt{NumPy} \citep{numpy},
\texttt{PyTorch} \citep{pytorch},
\texttt{pytorch-geometric} \citep{PyG},
\texttt{SciPy} \citep{scipy},
\texttt{shap} \citep{SHAP},
\texttt{statsmodels} \citep{statsmodels},
\texttt{treecorr} \citep{treecorr}
}

\begin{acknowledgements}
We thank the anonymous referee for their useful comments.
The authors are also grateful to Peter Behroozi, Haley Bowden, Suchetha Cooray, Daniela Galarraga-Espinosa, Martin Rey, Tjitske Starkenburg, Paco Villaescusa-Navarro, and Sandy Yuan, who provided valuable discussions and comments on this work.
We also thank the Kavli Institute for Theoretical Physics ``Building a Physical Understanding of Galaxy Evolution with Data-driven Astronomy'' program, where this work began.
This research was supported in part by the National Science Foundation under Grant No. NSF PHY-1748958.
This work used Rockfish - GPU resources at Johns Hopkins University through allocation PHY230101 from the Advanced Cyberinfrastructure Coordination Ecosystem: Services \& Support (ACCESS) program, which is supported by National Science Foundation grants NSF PHY-2138259, PHY-2138286, PHY-2138307, PHY-2137603, and PHY-2138296.
\end{acknowledgements}

\appendix
\section{GNN details} \label{app:GNN-details}

\subsection{Data preprocessing}

Model training is often easier if the inputs resemble (or are close to) Gaussian distributions.
We normalize all subhalo velocities by 100~km~s$^{-1}$, and logarithmically scale the subhalo properties $M_{\rm halo}$, $V_{\rm max}$, and $M_\star$.
We also rescale the DisPerSE cosmic web parameters so that they have zero mean and unit variance.
In order to ensure translational invariance, we also recenter all subhalo positions to the center of mass of the subhalos.

\subsection{GNN architecture details}

In Section~\ref{sssec:gnn-architecture}, we described the GNN architecture, which is composed of $N_{\rm unshared} = 8$ unique versions of the edge MLP $\bm\phi^{(\ell)}$, and a node MLP $\bm \psi$ that comprises three parts, $\bm \psi_1$, $\bm\psi_2$, and $\bm\psi_3$.
The network $\bm\psi_1$ ingests concatenated outputs of the edge MLPs, and $\bm\psi_2$ ingests the node features; $\bm\psi_3$ ingests the outputs of $\bm\psi_1$ and $\bm\psi_2$ and returns the final prediction.
Panel (c) of Figure~\ref{fig:gnn-schematic} also provides a high-level graphic of the GNN layers.
We use $N_{\rm hidden}=64$ and $N_{\rm latent} = 16$.
The total parameter count is 89,650: 50,304 for all $\{\bm\phi^{(\ell)}\}_{(\ell)}$, 26,976 for $\bm\psi_1$, 5,712 for $\bm\psi_2$, and 6,658 for $\bm\psi_3$.
In Table~\ref{tab:mlp-details}, we list the layer names, parameter counts, input sizes, and output sizes for all neural network layers in our GNN.

\begin{deluxetable*}{llrrr}
    \tablehead{\multicolumn{2}{l}{Layer} & \colhead{Number of parameters} & \colhead{Input size} & \colhead{Output size}
    }
    \startdata
    \multicolumn{2}{c}{$\bm\phi^{(\ell)}$ (Edge MLP)} & Total: 6,288 & 12 & 16 \\
    \hline
    &Linear1     & 832  & 12 & 64 \\
    &LayerNorm1  & 128  & 64 & 64 \\
    &SiLU1       & ---  & 64 & 64 \\
    &Linear2     &4,160 & 64 & 64 \\
    &LayerNorm2  & 128  & 64 & 64 \\
    &SiLU2       & ---  & 64 & 64 \\
    &Linear3     &1,040 & 64 & 16 \\
    \hline
    \multicolumn{2}{c}{$\bm\psi_1$ (Node MLP 1)} & Total: 26,976 & 384 & 16 \\
    \hline
    &Linear1     & 24,640 & 384 & 64 \\
    &LayerNorm1  & 128  & 64 & 64 \\
    &SiLU1       & ---  & 64 & 64 \\
    &Linear2     & 1,040 & 64 & 64 \\
    &LayerNorm2  & 128  & 64 & 64 \\
    &SiLU2       & ---  & 64 & 64 \\
    &Linear3     & 1,040 & 64 & 16 \\
    \hline
    \multicolumn{2}{c}{$\bm\psi_2$ (Node MLP 2)} & Total: 5,712 & 3 & 16 \\
    \hline
    &Linear1     & 256 & 3 & 64 \\
    &LayerNorm1  & 128  & 64 & 64 \\
    &SiLU1       & ---  & 64 & 64 \\
    &Linear2     & 4,160 & 64 & 64 \\
    &LayerNorm2  & 128  & 64 & 64 \\
    &SiLU2       & ---  & 64 & 64 \\
    &Linear3     & 1,040 & 64 & 16 \\ 
    \hline
    \multicolumn{2}{c}{$\bm\psi_3$ (Node MLP 3)} & Total: 6,658 & 32 & 2 \\
    \hline
    &Linear1     & 2,112 & 32 & 64 \\
    &LayerNorm1  & 128  & 64 & 64 \\
    &SiLU1       & ---  & 64 & 64 \\
    &Linear2     & 4,160 & 64 & 64 \\
    &LayerNorm2  & 128  & 64 & 64 \\
    &SiLU2       & ---  & 64 & 64 \\
    &Linear3     & 130 & 64 & ~2  
\tablecaption{Details for all MLPs in the GNN. The total parameter counts are given for each MLP, as well as the breakdown for each individual layer. The full model includes $N_{\rm unshared} = 8$ versions of $\bm\phi^{(\ell)}$. The total parameter count is 89,650.
\label{tab:mlp-details}}
\enddata
\end{deluxetable*}

We also tested an architecture variant with a single MLP $\bm\psi$ that takes in all $3 \times N_{\rm unshared} \times N_{\rm latent} = 768$ pooled edge features alongside the two node features, and directly makes predictions. 
However, there are many more pooled edge features than node features, which makes it difficult for the network to learn a robust combination in a single MLP.
We found that the GNN performance improves when we train $\bm\psi_1$ to distill the 768 (highly correlated) edge features into 16 features using $\bm\psi_1$.

\subsection{GNN optimization hyperparameters}

We performed a basic hyperparameter search over several variables.
The results are summarized below, along with more general remarks about GNN and neural network optimization. 
In particular, the first two choices---removing loops and using the \texttt{ClusterLoader} sampler---dramatically improved our GNN performance.

\textbf{Remove (self-)loops from graph.} Loops on graphs, i.e., edges that connect a node to itself, have a negative effect on performance.
This is likely because loops interfere with edge features $d_{ii}^{(\mathbf{x})} = 0$, which in turn makes it difficult for the model to learn a $1/d^n$ power law that describes gravity, tides, or other local interactions.

\textbf{Sample large batches of clustered nodes beyond the linking length.}
We found that the Pytorch-Geometric  \texttt{ClusterLoader} sampling strategy \citep{Chiang+2019} outperforms random node sampling or simple neighborhood-based sampling.
The \texttt{ClusterLoader} method uses graph clustering to identify relatively dense subgraphs and batch them together, resulting in efficient training and significantly lower loss.
We note that, while the \texttt{ClusterLoader} algorithm performs well for large and well-connected graphs, it seems to perform poorly at very small linking lengths (i.e. at $L \approx 0.3$~Mpc the GNN performs worse than the EBM models; Figure~\ref{fig:optimal-linking-length}).
We speculate that the \texttt{ClusterLoader} sampling strategy fails in the case of very sparse graphs, resulting in inefficient training.

\textbf{Adaptive optimization.} 
We use the Adam optimizer with decoupled weight decay (\texttt{AdamW}; \citealt{KingmaBa2014,loshchilov2018decoupled}).
We set the $(\beta_1, \beta_2)$ momentum parameters to $(0.9, 0.95)$.

\textbf{Learning rate = $\bm{10^{-2}}$.} 
The learning rate determines the size of the model parameter updates. 
This important hyperparameter also covaries with many other hyperparameters. 
Because we are using a relatively small neural network, we found that the learning rate can be made fairly large (i.e., relative to the values of $\sim 10^{-4}-10^{-5}$ often used for training deeper neural networks). 

\textbf{Weight decay~$\bm{=10^{-4}}$.} Weight decay is analogous to L2 regularization when using the \texttt{AdamW} optimizer. 
We found that weight decay lowers the gap between training and validation loss.

\textbf{Learning rate schedule.} 
Although we begin training at a relatively high learning rate ($0.01$), we reduce the learning rate to $0.002$ at epoch 75, $0.0004$ at epoch 150, and $0.00008$ at epoch 225.
This annealing process helps stabilize the optimization procedure and achieve a lower loss.

\textbf{Noise augmentation.} 
We add random noise to the node and edge features of the GNN to help it learn representations that are robust to noise \citep[e.g.,][]{Murphy+2019,Godwin+2022}.
During training, we sample and add Gaussian-distributed, zero-mean noise scaled to $0.0003$ times each input feature's scatter.

\bibliography{bib} %
\bibliographystyle{aasjournal}
\end{document}